\documentclass[useAMS,usenatbib]{mnras}
\usepackage{txfonts}
\usepackage{graphicx}
\usepackage[mathscr]{eucal}
\newcommand{\be}{\begin{equation}}
\newcommand{\ee}{\end{equation}}
\newcommand{\bdm}{\begin{displaymath}}
\newcommand{\edm}{\end{displaymath}}
\def\dmf{\dot{\mathfrak{M}}}
\def\msE{\mathscr{E}}
\title[Magnetic field of Be/X-ray pulsars in the SMC]
{On the magnetic fields of Be/X-ray pulsars in the Small Magellanic Cloud}
\author[N.R.\,Ikhsanov and S.\,Mereghetti]
{N.R.\,Ikhsanov$^{1,2,3}$\thanks{E-mail: n.ikhsanov@spbu.ru (NRI);
sandro@iasf-milano.inaf.it (SM)}
and S.\,Mereghetti$^4$ \\
$^1$ Pulkovo Observatory, Pulkovskoe shosse 65-1,  St. Petersburg,
196140, Russia\\
$^2$Special Astrophysical Observatory RAS, Nizhny Arkhyz, 369167
Russia\\
$^3$Saint-Petersburg State University, St.\,Petersburg, 198504 Russia\\
$^4$INAF, IASF-Milano, via E.Bassini 15, Milano I-20133, Italy}

\begin{document}
\date{Accepted: 2015 September 9; Received: 2015 July 8; Published: MNRAS {\bf
454}, 3760-3765 (2015)}
\pagerange{\pageref{firstpage}--\pageref{lastpage}} \pubyear{2015}
\maketitle
\label{firstpage}
 \begin{abstract}
We explore the possibility of explaining the properties of the
Be/X-ray pulsars observed in the Small Magellanic Cloud (SMC) within
the magnetic levitation accretion scenario. This implies that their
X-ray emission is powered by a wind-fed accretion on to a neutron
star (NS) which captures matter from a magnetized stellar wind. The
NS in this case is accreting matter from a non-Keplerian
magnetically levitating disc which is surrounding its magnetosphere.
This allows us to explain the observed periods of the pulsars in
terms of spin equilibrium without the need of invoking dipole
magnetic fields outside the usual range $\sim10^{11}- 10^{13}$\,G
inferred from cyclotron features of Galactic high-mass X-ray
binaries. We find that the equilibrium period of a NS, under certain
conditions, depends strongly on the magnetization of the stellar
wind of its massive companion and, correspondingly, on the magnetic
field of the massive companion itself. This may help to explain why
similar NSs in binaries with similar properties rotate with
different periods yielding a large scatter of periods of the
accretion-powered pulsar observed in SMC and our galaxy.
 \end{abstract}

\begin{keywords}
accretion, accretion discs -- pulsars: general -- stars: winds,
outflows -- X-rays: binaries.
\end{keywords}

\section{Introduction}

The Small Magellanic Cloud (SMC) contains a large number of
accreting pulsars in high mass  X-ray binaries (HMXBs). These
sources, being at a well known, virtually uniform distance and with
small interstellar absorption, constitute an ideal sample for
population studies  of neutron star (NS) binaries
\citep{Haberl-Pietsch-2004, Shtykovskiy-Gilfanov-2005,
Laycock-etal-2010}.

\citet{Klus-etal-2014} have recently reported parameters of 42
Be/X-ray binaries in the SMC observed   with the {\it Rossi X-ray
Timing Explorer} ({\it RXTE}) satellite over a time span of 14\,yr.
These systems, most of which are transients,  contain pulsars in
which the X-ray emission is powered by  wind-fed  accretion on to a
magnetized NS.  The average spin periods of these pulsars are in the
range $P_{\rm s} \sim 2.37 - 1323$\,s and change at an average rate
$|\dot{P}| \sim (0.02 - 620) \times 10^{-2}\,{\rm s\,yr^{-1}}$.
Their orbital periods range from about 4 to 500\,d.

\citet{Klus-etal-2014} estimated the magnetic fields of these NSs
considering the situation in which they accrete matter from a
Keplerian disc and  rotate close to their  equilibrium spin period. In this case,
the estimated fields of the majority of
the stars  (including all those with $P_{\rm s} > 100$\,s) turn out  to
be over the quantum critical value $B_{\rm cr} = m_{\rm e}^2 c^3/e \hbar \simeq
4.4 \times 10^{13}$\,G. Here $m_{\rm e}$ and $e$ are the mass and electric
charge of an electron, $c$ is the speed of light and $\hbar$ is the reduced Planck constant.
 Alternatively, if these NSs are not close to spin
equilibrium,   their inferred  magnetic fields are smaller than
$\sim10^{10}$\,G. Such results are rather unexpected since the
majority of NSs in our Galaxy, including accretion-powered pulsars
in HMXBs in which the   magnetic field is
measured from  the cyclotron resonance scattered features, have
magnetic fields between $10^{11}$ and $10^{13}$\,G \citep[see,
e.g.,][]{Revnivtsev-Mereghetti-2015}.  An attempt to invoke
currently used quasi-spherical accretion scenarios did not  help
much to improve the situation either, leading to estimated  magnetic
fields in excess of $10^{13}$\,G.

It is not unusual that the magnetic fields  evaluated from the spin
parameters of NSs in Galactic  HMXBs significantly exceed  those
measured through observations of their cyclotron resonance features.
As recently indicated by \citet{Ikhsanov-etal-2014}, this
inconsistency may reflect an oversimplification of currently used
wind-fed accretion scenarios, in which the magnetic field of the
matter captured by the NS from its environment is neglected. The
incorporation of the fossil magnetic field of the accreting matter
into the model leads, under certain conditions, to a different
accretion regime which is referred to as magnetic levitation
accretion. In this scenario, which is briefly outlined in the next
section,  accretion occurs through a non-Keplerian magnetically
levitating disc (ML-disc) and the maximum possible torque exerted on
the NS  significantly exceeds that previously evaluated in
traditional non-magnetic scenarios. This leads to a new expression
for the equilibrium period (see Section\ref{equilibrium}), which
allows us to explain the observed values of the spin and orbital
periods of the SMC pulsars without invoking magnetic fields outside
the canonical range of $10^{11}$--$10^{13}$\,G
(Section\,\ref{corbet}).

 \section{Magnetic levitation accretion}

We consider an HMXB, with orbital period $P_{\rm orb}$, composed of a
magnetized NS rotating with spin period $P_{\rm s}$ and a massive
early-type star, which underfills its Roche lobe and loses matter
through a stellar wind. The X-ray emission of the system is powered
by wind-fed accretion on to the NS. This implies that the NS
captures matter from the wind at a rate $\dmf \leq \dmf_{\rm c} =
\pi r_{\!_{\rm G}}^2 \rho_0 v_{\rm rel}$, where $r_{\!_{\rm G}} = 2
GM_{\rm ns}/v_{\rm rel}^2$ is the Bondi radius, $M_{\rm ns}$ is the
NS mass,  $v_{\rm rel}$  is its velocity  in the frame of the wind,
and $\rho_0 = \rho(r_{\!_{\rm G}})$ is the density of matter in the
region of interaction. The captured matter moves towards the NS
forming an accretion flow, which interacts with the stellar magnetic
field and confines the magnetosphere within the radius $r_{\rm m}$,
where the flow pressure is balanced by the pressure of the NS
magnetic field. The accreting matter penetrates into the field at
the magnetospheric boundary and, finally, falls on to the stellar
surface at the magnetic pole regions by moving along the magnetic
field lines.

The analysis of this accretion scenario reported by
\citet{Ikhsanov-Finger-2012} indicates that the structure of the
accretion flow, as well as the appearance of the accretion-powered
source, strongly depend on the physical conditions of the matter
captured at the Bondi radius, which, in  general, possesses some
angular momentum and magnetic field. Under these conditions, the
structure of the accretion flow beyond the magnetosphere can be
treated in the following basic approximations: (i)~a spherically
symmetrical or quasi-spherical flow, (ii)~a Keplerian disc, and
(iii)~an ML-disc. A key parameter
which allows us to determine which of these situations applies is
the relative velocity of the NS with respect to surrounding matter,
$v_{\rm rel}$.

The {\it spherically symmetrical} accretion occurs if the gas surrounding the
NS does not possess angular momentum and is non-magnetized. The captured matter
in this case moves towards the NS in the radial direction with the free-fall
velocity, $v_{\rm ff}(r) = \left(2GM_{\rm ns}/r\right)^{1/2}$ in a spherically
symmetrical fashion. Its density scales with the radius as $\rho(r) =
\dmf/\left(4 \pi r^2 v_{\rm ff}\right)$ and the ram pressure is
 \be
 \msE_{\rm ram}(r) = \rho(r) v_{\rm ff}^2(r) \propto r^{-5/2}.
 \ee
The minimum distance to which the spherical flow can approach a NS with
dipole magnetic moment $\mu$ is  $r \geq r_{\!_{\rm A}}$
\citep{Arons-Lea-1976, Elsner-Lamb-1977}, where
 \be\label{ra}
 r_{\!_{\rm A}} = \left(\frac{\mu^2}{\dmf (2GM_{\rm ns})^{1/2}}\right)^{2/7}
 \ee
is the Alfv\'en radius which is defined by equating the ram pressure
of the free-falling gas, with the magnetic pressure due to dipole
magnetic field of the NS, $p_{\rm m} = \mu^2/\left(2\pi r^6\right)
\propto r^{-6}$.

The structure of a non-magnetized accretion flow deviates from
spherically symmetrical if the captured matter possesses angular
momentum. The process of mass accretion in this case is accompanied
by the accretion of angular momentum, which in a binary system with
the orbital angular velocity $\Omega_{\rm orb} = 2 \pi/P_{\rm orb}$
occurs at the rate $\dot{J} = \xi\,\Omega_{\rm orb}\,r_{\!_{\rm
G}}^2\,\dmf$ \citep{Illarionov-Sunyaev-1975}. Here $\xi$ is the
parameter accounting for dissipation of angular momentum due to
density and velocity gradients in the accreting non-magnetized gas
\citep[see e.g.][and references therein]{Ruffert-1999}. The angular
velocity of matter in this so called {\it quasi-spherical
accretion flow} scales with the radius as
 \be
\Omega_{\rm f}(r) = \xi\,\Omega_{\rm orb}\,
 \left(\frac{r_{\!_{\rm G}}}{r} \right)^2,
 \ee
and the azimuthal component of the dynamical pressure of the flow,
 \be
 \msE_{\phi}(r) = \rho_0\,\xi^2\,\Omega_{\rm orb}^2\,r_{\!_{\rm G}}^2\,
 \left(\frac{r_{\!_{\rm G}}}{r}\right)^{7/2}\ \propto r^{-7/2},
 \ee
increases more rapidly than the ram pressure. The accretion proceeds
in a quasi-spherical fashion up to the circularization radius
  \be
r_{\rm circ} = \frac{\xi^2\,\Omega_{\rm orb}^2\,r_{\!_{\rm
G}}^4}{GM_{\rm ns}},
 \ee
at which condition $\msE_{\phi}(r) = \msE_{\rm ram}(r)$ is
satisfied. The angular velocity of the accreting matter,
$\Omega_{\rm f}(r)$, at this radius reaches the Keplerian angular
velocity, $\Omega_{\rm k} = \left(r^3/2GM_{\rm ns}\right)^{1/2}$,
and the accretion flow switches into a {\it Keplerian accretion
disc} \citep{Pringle-Rees-1972, Shakura-Sunyaev-1973}.

The accretion picture may differ from that presented above if the matter
captured by the NS at the Bondi radius is magnetized. As long as the Alfv\'en
velocity, $v_{\rm a}(r) = B_{\rm f}(r)/\left[4 \pi \rho(r)\right]^{1/2}$, in
the accreting matter is smaller than the free-fall velocity, the magnetic field
of the accreting gas, $B_{\rm f}$, does not influence significantly the flow
structure. Therefore, the captured matter initially follows ballistic
trajectories forming a quasi-spherical accretion flow in which the angular
momentum and magnetic flux are almost conserved. However, the magnetic pressure
in the free-falling gas increases rapidly,
 \be
 \msE_{\rm m}(r) = \frac{B_{\rm f}^2(r)}{8 \pi} \propto r^{-4},
 \ee
and reaches the ram pressure at the Shvartsman radius
\citep{Shvartsman-1971}
 \be\label{rsh}
 R_{\rm sh} = \beta_0^{-2/3}\ r_{\!_{\rm G}}\
   \left(\frac{c_{\rm so}}{v_{\rm rel}}\right)^{4/3},
 \ee
which is defined by equating the magnetic to ram pressure. Here
$\beta_0 = \beta(r_{\!_{\rm G}}) = 8 \pi \rho_0 c_{\rm so}^2/B_{\rm
f0}^2$ is the ratio of thermal to magnetic pressure in the captured
matter at the Bondi radius, $c_{\rm so} = c_{\rm s}(r_{\!_{\rm G}})$
is the sound speed and $B_{\rm f0} = B_{\rm f}(r_{\!_{\rm G}})$ is
the magnetic field in the accreting matter. The flow at this radius
is converted into a non-Keplerian ML-disc in which the accreting matter
is confined by its own magnetic field \citep[for discussion
see][]{Bisnovatyi-Kogan-Ruzmaikin-1976, Igumenshchev-etal-2003}. The
accretion inside the ML-disc proceeds on the time-scale of the
magnetic flux dissipation and, in the general case, can be treated
in a diffusion approximation. The value of the effective diffusion
coefficient strongly depends on the configuration of the magnetic
field in the disc and its stability. It ranges from the Bohm
diffusion coefficient, if the magnetic field annihilation in the
disc is governed by dissipative instabilities and magnetic
reconnections, up to much larger values if the field configuration
is interchange unstable \citep[for discussion see][and references
therein]{Igumenshchev-2009, Tchekhovskoy-etal-2011,
Dexter-etal-2014}.

Thus, a {\it quasi-spherical} accretion on to a NS in a wind-fed
HMXB occurs if the Alfv\'en radius is larger than both the
circularization and the Shvartsman radii. The angular momentum and
magnetic field of the accreting material in this case are too small
to significantly influence the flow structure before the ballistic
trajectories of the accreting matter are truncated by the magnetic
field of the NS. Solving the inequality $r_{\!_{\rm A}} \geq
\max\{r_{\rm circ}, R_{\rm sh}\}$ for $v_{\rm rel}$ one finds
    \be
v_{\rm rel} \geq \left\{
\begin{array}{lc}
v_{\rm cr}, \hspace{-1mm}& {\rm for} \hspace{1mm} R_{\rm
sh} < r_{\rm circ},\\
& \\
v_{\rm ma}, \hspace{-1mm}& {\rm for} \hspace{1mm} R_{\rm sh} >
r_{\rm circ},
 \end{array}
\right.
   \ee
where \citep{Ikhsanov-2007}
 \be
 v_{\rm cr} \simeq 160\,{\rm km\,s^{-1}} \xi_{0.2}^{1/4} \mu_{30}^{-1/14} m^{11/28}
\dmf_{15}^{1/28} P_{50}^{-1/4},
 \ee
and \citep{Ikhsanov-Beskrovnaya-2012, Ikhsanov-Finger-2012}
   \be\label{vma}
 v_{\rm ma} \simeq 460\,{\rm km\,s^{-1}}\ \times\ \beta_0^{-1/5}\
 \mu_{30}^{-6/35}\ m^{12/35}\ \dmf_{15}^{3/35}\
 \left(\frac{c_{\rm so}}{10\,{\rm km\,s^{-1}}}\right)^{2/5}.
 \ee
Here $\mu_{30} = \mu/\left(10^{30}\,{\rm G\,cm^3}\right)$ is the
dipole magnetic moment ($\mu = (1/2) B_{\rm ns} R_{\rm ns}^3$) of a
NS with surface magnetic field $B_{\rm ns}$ and radius $R_{\rm ns}$,
$m = M_{\rm ms}/1.4$\,M$_{\sun}$, $\dmf_{15} =
\dmf/\left(10^{15}\,{\rm g\,s^{-1}}\right)$, $P_{50} = P_{\rm
orb}/50\,{\rm d}$, and the parameter $\xi_{0.2} = \xi/0.2$ is normalized
according to \citet{Ruffert-1999}. The radius of the magnetosphere,
within this scenario, is comparable or slightly exceeds the
Alfv\'en radius \citep{Arons-Lea-1976, Elsner-Lamb-1977}, and
accretion beyond the magnetospheric boundary proceeds in the form of
a free-falling gas or a hot turbulent spherical envelope
\citep{Lamb-etal-1977, Ikhsanov-2001, Ikhsanov-2003}.

A {\it Keplerian accretion disc} in a wind-fed HMXB can form if the
circularization radius exceeds both the Alfv\'en and Shvartsman,
radii. Solving inequality $r_{\rm circ} \geq \max\{r_{\!_{\rm A}},
R_{\rm sh}\}$ for $v_{\rm rel}$ one finds
    \be
v_{\rm rel} \leq \left\{
\begin{array}{lc}
v_{\rm cr}, \hspace{-1mm}& {\rm for} \hspace{1mm} R_{\rm
sh} < r_{\!_{\rm A}},\\
& \\
v_{\rm ca}, \hspace{-1mm}& {\rm for} \hspace{1mm} R_{\rm sh} >
r_{\!_{\rm A}},
 \end{array}
\right.
   \ee
where \citep{Ikhsanov-etal-2015}
 \be
v_{\rm ca} \simeq 80\,{\rm km\,s^{-1}}
\xi_{0.2}^{3/7}\,\beta_0^{1/7}\,m^{3/7}\,P_{50}^{-3/7}
 \left(\frac{c_{\rm so}}{10\,{\rm km\,s^{-1}}}\right)^{-2/7}.
 \ee

Finally, a {\it non-Keplerian ML-disc} can form in a wind-fed HMXB
if the Shvartsman radius is larger than both, the Alfv\'en and
circularization radii. Solving inequality $R_{\rm sh} \geq
\max\{r_{\!_{\rm A}}, r_{\rm circ}\}$ for $v_{\rm rel}$ yields
 \be
v_{\rm ca} < v_{\rm rel} < v_{\rm ma}.
 \ee
The inequality $v_{\rm ca} < v_{\rm ma}$ is satisfied if $\beta_0 <
\beta_{\rm max}$, where
 \be\label{betamax}
 \beta_{\rm max} \simeq 164\ \xi_{0.2}^{-5/4}\ m^{-1/4}\ P_{50}^{5/4}\
 \mu_{30}^{-1/2}\ \dmf_{15}^{1/4}
 \left(\frac{c_{\rm so}}{10\,{\rm km\,s^{-1}}}\right)^2.
 \ee
This indicates that the {\it magnetic levitation accretion} scenario
occurs if the fossil magnetic field in the matter captured by the NS
is $B_{\rm f}(r_{\!_{\rm G}}) \geq B_{\rm min}$, where
 \begin{eqnarray}\label{bmin}
 B_{\rm min} & = & \left(\frac{2\,\dmf\,v_{\rm rel}^3\,c_{\rm so}^2}
 {(GM_{\rm ns})^2\,\beta_{\rm max}}\right)^{1/2}
 \simeq 6 \times 10^{-4}\,{\rm G}\ \times \\
   \nonumber
  & \times & \xi_{0.2}^{5/8}\ \mu_{30}^{1/4} m^{-7/8}\ P_{50}^{-5/8}\
  \dmf_{15}^{3/8}
   \left(\frac{v_{\rm rel}}{100\,{\rm km\,s^{-1}}}\right)^{3/2}.
  \end{eqnarray}

The matter in an ML-disc approaches the NS until the radius
  \be\label{rma}
 r_{\rm ma} = \left(\frac{c\,m_{\rm p}^2}{16\,\sqrt{2}\,e\,k_{\rm B}}\right)^{2/13}
 \frac{\alpha_{\rm B}^{2/13} \mu^{6/13} (GM_{\rm ns})^{1/13}}{T_0^{2/13}
 \dmf^{4/13}},
 \ee
where the pressure exerted by the disc on to the stellar
magnetosphere is equal to the magnetic pressure due to the NS dipole
field and the rate of diffusion of the accreting matter into the
stellar field is equal to the mass capture rate by the star from its
environment \citep[for discussion see][]{Ikhsanov-2012,
Ikhsanov-etal-2014}. Here $m_{\rm p}$ is the proton mass, $k_{\rm
B}$ is the Boltzmann constant and $T_0$ is the temperature of the
matter at the inner radius of the disc. The parameter
$\alpha_{\!_{\rm B}} = D_{\rm eff}/D_{\rm B}$ is the ratio of the
effective coefficient of diffusion of the accreting matter into the
stellar field at the magnetospheric boundary, $D_{\rm eff}$, to the
Bohm diffusion coefficient, which in the considered case can be
expressed as $D_{\rm B} = c k_{\rm B} T_0 r_{\rm ma}^3/(32 e \mu)$.
Numerical simulations of the diffusion process and measurements of
the rate at which the solar wind penetrates into the magnetosphere
of the Earth suggest that $\alpha_{\!_{\rm B}} \sim$\,0.01--1
\citep{Gosling-etal-1991}. The matter being diffused into the
stellar field flows along the magnetospheric field lines and reaches
the  NS surface  at the magnetic pole regions.

The possibility of explaining the parameters of the SMC Be/X-ray
pulsars within the quasi-spherical and the Keplerian disc accretion
scenarios has been already discussed by \citet{Klus-etal-2014}. Here
we explore the possibility that these pulsars accrete matter from an
ML-disc. In the next section we evaluate the equilibrium period of an
NS in this scenario and present the expected Corbet ($P_{\rm s}$ versus
$P_{\rm orb}$) diagram in Section\,\ref{corbet}.

 \section{Equilibrium period}\label{equilibrium}

The equation governing the spin evolution of an NS accreting matter from
an ML-disc reads
 \be\label{spin-evol}
 2 \pi I \dot{\nu} = K_{\rm a} + K_{\rm b} + K_{\rm c},
 \ee
where $I$ is the moment of inertia and $\nu = 1/P_{\rm s} = \omega_{\rm s}/2\pi$ is the spin
frequency of the NS. The first term in the right-hand side of this
equation,
 \be\label{k-acc}
 K_{\rm a} = \dmf\ \ell(r_{\rm ma}) \simeq \dmf\,\omega_{\rm s}\,r_{\rm ma}^2,
 \ee
is the rate at which  angular momentum is transferred to the NS by
the matter flowing inside the magnetosphere. Here $\ell(r_{\rm ma})
\simeq r_{\rm ma} \times (r_{\rm ma} \omega_{\rm s})$ is the
specific angular momentum of matter in the magnetopause at the
rotational equator. This matter, which penetrates into the stellar
field at the magnetospheric boundary, corotates with the NS and
flows towards its surface along the field lines. $K_{\rm a}$ has a
positive sign and represents the minimum possible spin-up torque
exerted on to a star which is surrounded by a magnetosphere of
radius $r_{\rm ma}$ and accretes matter on to its surface at a rate
$\dmf$.

The second term \citep{Ikhsanov-2012, Ikhsanov-etal-2014},
 \be\label{k-mag}
 K_{\rm b} = \frac{k_{\rm t}\,\mu^2}{\left(r_{\rm ma}\,r_{\rm
 cor}\right)^{3/2}}\ \left(\frac{\Omega_{\rm f}(r_{\rm ma})}{\omega_{\rm
 s}} - 1\right),
 \ee
accounts for the angular momentum exchanged between the star and the
disc at the magnetospheric boundary  (here $k_{\rm  t}$ is a
dimensionless parameter of order unity). The sign of this term
depends on the ratio between the angular velocity of matter at the
inner radius of the disc, $\Omega_{\rm f}(r_{\rm ma})$, and the
angular velocity of the NS.

The last term, $K_{\rm c}$,  accounts for the angular momentum
exchanged between the NS and the disc at radii $r > r_{\rm ma}$.
Under the conditions of interest (d$\Omega_{\rm f}/$d$r \leq 0$), this
term is smaller than $K_{\rm b}$ at least by a factor  $\sim
\left(r/r_{\rm ma}\right)^{3/2}$. Therefore, its contribution is
relatively small and we neglect it in the following.

The equilibrium period, $P_{\rm eq}$, is defined by equating the
total torque exerted on the NS (right-hand side of
equation~\ref{spin-evol}) to zero, which, using  equations~(\ref{k-acc}) and
(\ref{k-mag}), yields
 \be
  P_{\rm eq} \simeq  P_{\rm f}(r_{\rm ma})
 \left[1 - \frac{1}{\sqrt{2}k_{\rm t}} \left(\frac{r_{\rm ma}}{r_{\!_{\rm
 A}}}\right)^{7/2}\right],
 \ee
where $P_{\rm f}(r_{\rm ma}) = 2 \pi/\Omega_{\rm f}(r_{\rm ma})$.

This equation suggests that the angular velocity of the NS in spin equilibrium
exceeds the angular velocity of the matter at the inner radius of the disc by a
factor of $\left[1 - (1/\sqrt{2}k_{\rm t}) \left(r_{\rm ma}/r_{\!_{\rm
A}}\right)^{7/2}\right]$. The torque $K_{\rm b}$ in this case has a negative
sign and tends to spin-down the star. It, however, is compensated by the
spin-up torque $K_{\rm a}$ and thus, the total toque exerted on the NS is zero.

Evaluating the ratio
 \be
\left(\frac{r_{\rm ma}}{r_{\!_{\rm
 A}}}\right)^{7/2} \sim 10^{-3}\
 \alpha_{\rm B}^{7/13}\ \mu_{30}^{-5/13}\ m^{10/13}\
 \dmf_{15}^{-1/13} T_6^{-7/13},
 \ee
one finds that the expression in the brackets in the conditions of interest
($10^{-3} \ll k_{\rm t} \leq 1$) is close to unity \citep[here $T_6 =
T_0/10^6$\,K is normalized to the temperature of optically thick gas irradiated
by the pulsar emission, see, e.g.][]{Hickox-etal-2004}. This indicates that the
angular velocity of the NS in spin equilibrium is close to the angular velocity
of the matter at the inner radius of the disc. This allows us for simplicity to
approximate the equilibrium period of the NS by $P_{\rm f}(r_{\rm ma})$.

The analysis of the transport of angular momentum across the ML-disc
is rather complicated and beyond the scope of this paper. The
angular momentum extracted from the NS can be stored in the disc
\citep{Sunyaev-Shakura-1977, D-Angelo-Spruit-2012} or/and
transferred out from the disc by Alfv\'en waves. Here we simply
assume that the angular velocity of the matter in the disc scales
with the radius as $\Omega_{\rm f}(r) \propto (R_{\rm
sh}/r)^{\gamma}$, where $\gamma$ is a free parameter, which in the
general case is limited to\footnote{The value $\gamma$=2 corresponds
to conservation of angular momentum, $\gamma$=1.5 to Keplerian
rotation, $\gamma$=1 to rotation with constant linear velocity, and
$\gamma$=0 to solid body rotation.} $\gamma \leq 2$. The angular
velocity of matter at the inner radius of the disc can be expressed
as
 \be\label{omfma}
 \Omega_{\rm f}(r_{\rm ma}) = \Omega_{\rm f}(R_{\rm sh})
 \left(\frac{R_{\rm sh}}{r_{\rm
  ma}}\right)^{\gamma},
 \ee
where
 \be\label{omfsh}
 \Omega_{\rm f}(R_{\rm sh}) \simeq
 \xi\,\Omega_{\rm orb} \left(\frac{r_{\!_{\rm G}}}{R_{\rm sh}}\right)^2
 \ee
is the angular velocity of matter at the Shvartsman radius. This is
derived by taking into account that the magnetic field of the
accreting matter does not significantly influence the flow structure
beyond the Shvartsman radius and the matter in the region $R_{\rm
sh} < r\leq r_{\!_{\rm G}}$ moves towards the NS quasi-spherically.

Putting equations~(\ref{rsh}), (\ref{rma}) and (\ref{omfsh}) into equation~(\ref{omfma})
and taking into account that the relative velocity in the considered case is
$v_{\rm rel} = \zeta v_{\rm ma}$,  with $\zeta<1$, one finds
 \be\label{pfsh}
  P_{\rm f}(r_{\rm ma}) = 0.5 \left(\frac{1.3^{10\gamma/3}}{2^{\gamma}}\right)\
   A_{\rm m} P_{\rm orb},
 \ee
where
  \be
  A_{\rm m} = \xi^{-1} \beta_0^{-4/5} c_{\rm so}^{8/5}
   \mu^{a_1} \dmf^{-a_2} \zeta^{a_3} (GM_{\rm ns})^{a_4}
   \left(\frac{c\,m_{\rm p}^2 \alpha_{\!_{\rm B}}}{16 \sqrt{2} e k_{\rm B} T_0}
 \right)^{2\gamma/13}
  \ee
and
 \bdm
 a_1 = (208 - 50\,\gamma)/455,
 \edm
 \bdm
 a_2 = (10\,\gamma +104)/455,
 \edm
 \bdm
 a_3 = (10\,\gamma - 8)/3,
 \edm
 \bdm
 a_4 = (100\,\gamma - 416)/455.
 \edm

Finally, the value of the dimensionless parameter $\zeta$ in the
general case lies in the interval $\zeta_{\rm min} < \zeta < 1$,
where
   \be\label{zeta}
\zeta_{\rm min} = \left\{
\begin{array}{lc}
v_{\rm ns}/v_{\rm ma}, \hspace{1mm}& {\rm for} \hspace{5mm}
v_{\rm ns} > v_{\rm ca}, \\
& \\
v_{\rm ca}/v_{\rm ma}, \hspace{1mm}& {\rm for} \hspace{5mm}
v_{\rm ns} < v_{\rm ca},
\end{array}
\right.
   \ee
and $v_{\rm ns}$ is the orbital velocity of the NS.

 \begin{figure*}
 \begin{center}
\includegraphics[width=15.cm]{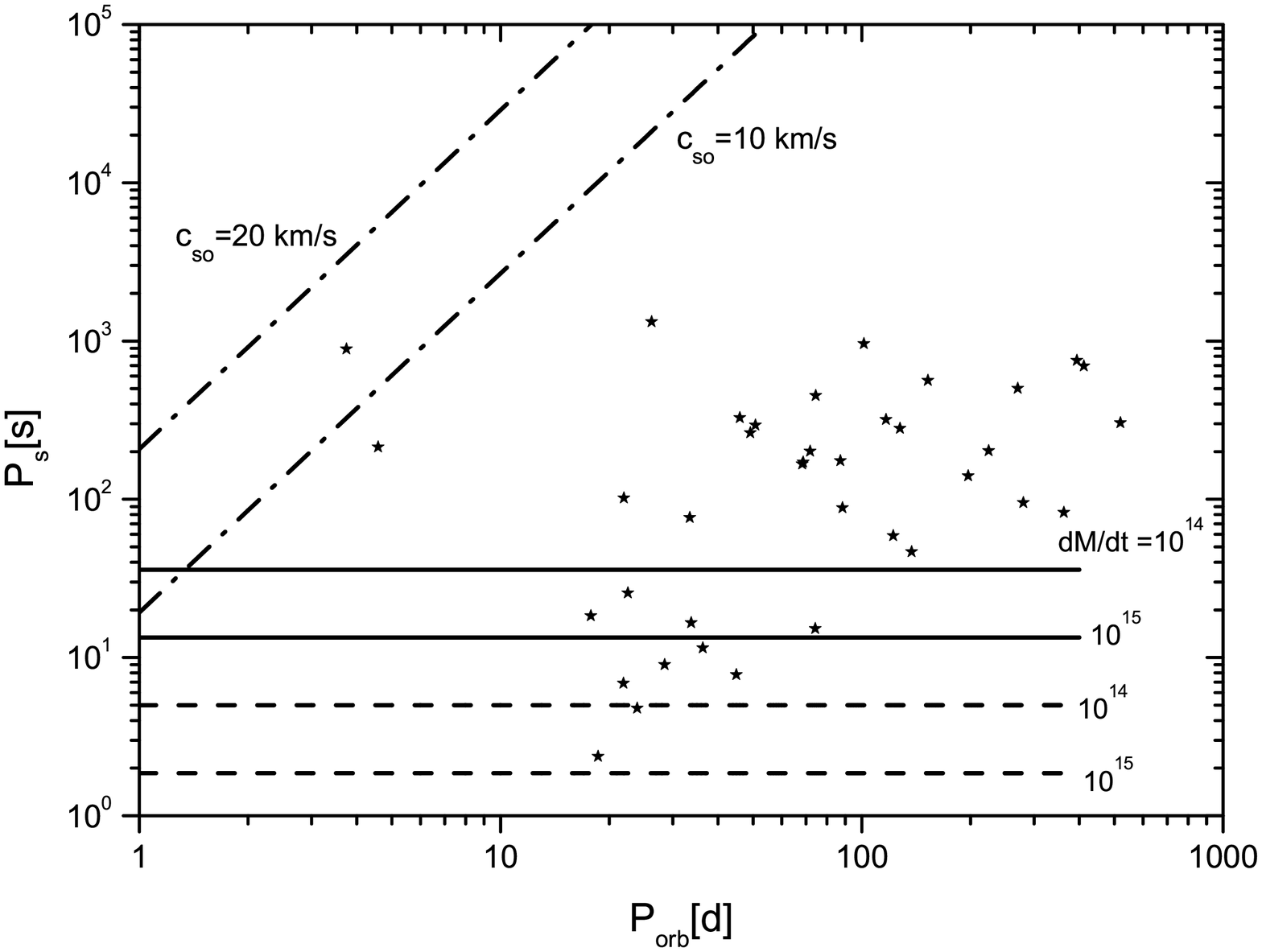}
 \caption{Corbet diagram of the Be/NS pulsars in the SMC,  with lines
of maximum and minimum equilibrium period in the ML-disc accretion
hypothesis. The solid lines are the minimum equilibrium period for
$\mu = 10^{30}$ G cm$^3$ and $\dmf =10^{14}$ and 10$^{15}$ g
s$^{-1}$. The dashed lines are the minimum equilibrium period for
$\mu = 10^{29}$ G cm$^3$ and $\dmf =10^{14}$ and 10$^{15}$ g
s$^{-1}$. The dash-dotted line indicates the maximum equilibrium
period (independent of $\mu$ and $\dmf$). }
 \label{f1}
 \end{center}
 \end{figure*}

 \section{Application to the SMC pulsars}\label{corbet}

In this section we explore the possibility of explaining the spin
properties of the SMC Be/X-ray pulsars in terms of the magnetic
levitation accretion scenario described above. We assume that the
NSs in these systems rotate close to the equilibrium period $P_{\rm
eq} \simeq P_{\rm f}(r_{\rm ma})$ given by equation~(\ref{pfsh})  and
accrete matter from an ML-disc. This implies that the relative
velocity of the NSs in the frame of stellar wind of their companions
meets the condition $v_{\rm cr} \leq v_{\rm rel} \leq v_{\rm ma}$
and $\beta_0 < \beta_{\rm max}$ expressed by equation~(\ref{betamax}).
Since the accretion process in an ML-disc is fully controlled by the
magnetic field of the accreting matter itself, we consider the case
of rigid rotation of the disc by setting $\gamma = 0$.

The minimum possible value of the equilibrium period, $P_{\rm min}$,
which an NS can acquire within the magnetic levitation accretion
scenario can be derived from equation~(\ref{pfsh}), by setting $\beta_0 =
\beta_{\rm max}$ and, correspondingly, $\zeta = 1$. This yields
    \be\label{pmin}
P_{\rm min}^{\rm eq} \simeq 14\,{\rm s} \times
 \mu_{30}^{6/7}\ \dmf_{15}^{-3/7}\ m^{-5/7},
%\mu_{30}^{0.86}\ \dmf_{15}^{-0.43}\ m^{-0.71},
  \ee
where $\dmf_{15}$ is the average mass accretion rate normalized to
$10^{15}\,{\rm g\,s^{-1}}$.

In Fig.\,\ref{f1} the values of $P_{\rm min}^{\rm eq}$ are compared
to the observed values of $P_{\rm s}$ and $P_{\rm orb}$ for
representative values of  $\mu$ (10$^{30}$ G cm$^3$ solid lines;
10$^{29}$ G cm$^3$ dashed lines) and  $\dmf$ (10$^{14}$ and
10$^{15}$ g  s$^{-1}$). The corresponding lines indicates the
minimum possible equilibrium period of an NS which captures matter
from a relatively slow, moderately magnetized stellar wind in the
magnetic levitation accretion scenario.

The maximum possible equilibrium period, $P_{\rm max}$, can be
evaluated from equation~(\ref{pfsh}) by setting $\beta_0 = 1$ and $\zeta
= \zeta_{\rm min}$, which for the case $v_{\rm ns} < v_{\rm
ca}$ is (see equation~\ref{zeta}):
 \be
 \zeta_{\rm min} \simeq 0.17\ \beta_0^{12/35}\ \mu_{30}^{6/35}\
 \dmf_{15}^{-3/35}\ \xi_{0.2}^{3/7}\ m^{3/35}\ c_6^{-24/35}\ P_{50}^{-3/7}.
 \ee
Putting this value into equation~(\ref{pfsh}) one finds
  \be\label{pmax}
 P_{\rm max}^{\rm eq} \simeq 20\,{\rm s} \times
 P_{\rm orb(d)}^{15/7}\ \xi_{0.2}^{-15/7}\ \beta_0^{-12/7}\ c_6^{24/7}\ m^{-8/7}.
% P_{\rm orb(d)}^{15/7}\ \xi_{0.2}^{-15/7}\ \beta_0^{-1.71}\
% c_6^{3.43}\ m^{-1.14}.
 \ee
The function $P_{\rm max}^{\rm eq} = P_{\rm max}^{\rm eq}(P_{\rm
orb})$ is shown by the dot-dashed line in Fig.\,\ref{f1}. The
maximum equilibrium period of an NS which accretes from a magnetized
slow wind does not depend on the magnetic field of the NS itself and
on the X-ray luminosity of the pulsar. However, it is a strong
function of the sound speed in the surrounding matter and,
therefore, it can exceed the value given by equation~(\ref{pmax}) if the
temperature of the stellar wind exceeds $10^4$\,K.

Fig.\,\ref{f1} shows that practically all the pulsars reported by
\citet{Klus-etal-2014} have spin-period values smaller than  $P_{\rm
max} = P_{\rm max}(P_{\rm orb})$ and  lying above the lines of
$P_{\rm min}$ corresponding to $\mu\sim10^{\rm 29-30}$ G cm$^3$ and
accretion rates consistent with the long-term average luminosities
of these sources. This suggests that the observed spin periods can
be explained within the magnetic levitation accretion scenario with
surface magnetic fields of the NSs in the canonical interval $B_{\rm
ns}\sim10^{11}$--$10^{13}$\,G.

 \section{Discussion}\label{discussion}

We find that the observed spin periods of the SMC Be/X-ray pulsars
are in a range consistent with the values expected for the
equilibrium periods of NSs with magnetic fields of $B \sim
10^{\rm 11-13}$\,G and accreting from an ML-disk. The assumption that the
NSs in these systems are spinning close to an equilibrium value is
supported by the fact that they show alternate episodes of spin-up
and spin-down, which do not change $P_{\rm s}$ significantly on the
long term. We believe that the equilibrium period is set by the
average accretion rate experienced by these NS during the long time
intervals of quiescence (or low X-ray luminosity) between their
bright outbursts. In fact these sources spend most of the time in
such low-$\dmf$ conditions, resulting in X-ray luminosities well
below the {\it RXTE} sensitivity limit of $\sim10^{36}\,{\rm
erg\,s^{-1}}$. For this reason we have adopted in Fig.\,\ref{f1}
values of $\dmf$ corresponding to luminosities of
$\sim10^{34-35}\,{\rm erg\,s^{-1}}$.

The equilibrium period of an NS which accretes from a ML-disc tends
to increase with the orbital period of a binary system (see
equation~\ref{pfsh}). This, in particular, can be a reason for a lack of
pulsars in the lower-right part of the Corbet plot. On the other
hand, $P_{\rm eq}$ also depends on several other parameters, such as
the magnetic field and relative velocity of an NS, the mass accretion
rate, and physical conditions in the stellar wind with which the NS
interacts. The great diversity of possible combinations of these
parameters is responsible for the large scatter of the observed spin
periods in the Corbet plot. In this accretion regime, the magnetic
field of the massive star plays an important role, with stronger
fields leading to longer equilibrium periods of the pulsar.

Spectropolarimetric observations of O/B-type stars give evidence for
a relatively strong magnetization of these objects \citep[see,
e.g.][and references therein]{Walder-etal-2012}. The strength of the
large-scale field at the surface of several of these objects has
been measured in the range $\sim 500-5000$\,G, and in some cases
beyond 10\,kG \citep{Hubrig-etal-2006, Martins-etal-2010,
Oksala-etal-2010}. Some of the early type stars are surrounded by
X-ray coronae which indicate the magnetic activity of these objects
\citep{Schulz-etal-2003}. As most of these stars rotate relatively
fast \citep[see e.g.][and references therein]{Rosen-etal-2012}, the
magnetic field in the wind is dominated by the toroidal component
which scales with the radius $\propto r^{-1}$.

A similar situation is realized in the solar wind in which the
magnetic field at a distance of 1\,au is $B_{\rm sw} \sim
10^{-5}$\,G and the parameter $\beta$ is close to the equipartition
value, $\beta \sim 1$ \citep{Mullan-Smith-2006}. Following this
similarity one can suggest that the surface large-scale magnetic
field of massive stars in the considered systems is a factor of
$B_{\rm min}/B_{\rm sw} \sim 50$ larger than the surface large-scale
magnetic field of the Sun and can be as large as a few hundred
Gauss. A smaller magnetization of these stars cannot be also
excluded if the dynamo action applies in their outflowing discs. The
fact that periods of the considered pulsars are much shorter than
the maximum possible period predicted by our model may indicate that
the average value of $\beta_0$ exceeds unity and hence, the stellar
wind of early spectral type stars is less magnetized than the solar
wind. The observed range of periods of the pulsars ($\sim
1-1000$\,s) in this case can be explained in terms of variation of
$\beta_0$ parameter from system to system within an order of
magnitude.

Our study confirms the conclusion of \citet{Klus-etal-2014} that all
of the considered pulsars are situated in a relatively slow wind.
This is consistent with current views on the mass outflow process of
Be stars, in which the stellar wind at the equatorial plane is
dominated by a dense outflowing disc. The radial velocity of matter
in the disc is comparable or even smaller than the orbital velocity
of the NS \citep{Okazaki-Negueruela-2001}.

We finally note that our results apply also to the Be/X-ray pulsars
in our Galaxy, which show a distribution in the $P_{\rm orb}$-$P_{\rm
s}$ similar to that of the SMC sources. This supports the view that
the properties and evolution of HMXBs in the SMC and in our Galaxy
share a common nature and are governed by similar physical
processes.

\section*{Acknowledgements}

We would like to thank anonymous referee for very useful and
stimulating comments. NRI thanks INAF at Milano for kind hospitality
and acknowledges support of the Russian Scientific Foundation under
the grant no.\,14-50-00043. This work has been partially supported
through financial contribution from the agreement ASI/INAF
I/037/12/0 and from PRIN INAF 2014.

\bsp

\label{lastpage}

\end{document}